\documentclass[aps, pre, amsmath, amssymb, amsfonts, twocolumn, showpacs, floatfix]{revtex4}
\usepackage{graphicx}
\usepackage{subfigure}
\usepackage{dcolumn}
\usepackage{bm}
\usepackage{pst-node}
\usepackage{psfrag}
\usepackage{epsfig}

\newcommand     {\beq}[1]         { \begin{equation} #1 \end{equation} }
\newcommand     {\beqa}[1]        { \begin{eqnarray} #1 \end{eqnarray} }

\newcommand     {\EP}             { \varepsilon }
\newcommand     {\SI}             { \sigma }
\newcommand     {\EC}             { \varepsilon_c }
\newcommand     {\SC}             { \sigma_c }
\newcommand     {\AL}             { \alpha }

\newcommand     {\PPE}             { P(\varepsilon) }

\begin{document}

\title{Failure process of a bundle of plastic fibers}

\author{ Frank Raischel${}^1\footnote{Electronic 
address:raischel@ica1.uni-stuttgart.de}$, Ferenc Kun${}^{2}$, and Hans
J.\ Herrmann${}^{1,3}$}  
\affiliation{
${}^1$ ICP, University of Stuttgart, Pfaffenwaldring 27, D-70569
Stuttgart, Germany\\ 
${}^2$Department of Theoretical Physics, University of Debrecen, P.\
O.\ Box:5, H-4010 Debrecen, Hungary \\
${}^3$ Departamento de F\'\i sica, Universidade Federal do Cear\'a, Campus
do Pici, 60451-970 Fortaleza CE, Brazil}  

\date{\today}
\begin{abstract}
We present an extension of fiber bundle models considering
that failed fibers still carry a fraction $0 \leq \alpha \leq 1$ of their
failure load. The value of $\alpha$ interpolates between the perfectly
brittle failure $(\alpha = 0)$ and perfectly plastic behavior
$(\alpha=1)$ of fibers. We show that the finite load bearing capacity
of broken fibers has a substantial effect on the failure process of
the bundle. In the case of
global load sharing it is found that for $\alpha \to 1$ the
macroscopic response of the bundle becomes perfectly plastic with a 
yield stress equal to the average fiber strength. 
On the microlevel, the size distribution of avalanches has a crossover
from a power law of exponent $\approx 2.5$ to a faster exponential decay. For
localized load sharing, computer simulations revealed a sharp
transition at a well defined value $\alpha_c$ from a phase where
macroscopic failure occurs due to localization as a consequence of
local stress enhancements, to another one where the disordered fiber
strength dominates the damage process. Analysing the microstructure of
damage, the transition proved to be  analogous to
percolation. At the critical point $\alpha_c$, the spanning cluster
of damage is found to be compact with a fractal boundary.
The distribution of bursts of fiber breakings shows a power law
behaviour with a universal exponent $\approx 1.5$ equal to  the mean field exponent
of fiber bundles of critical strength distributions. The model can be relevant to understand the shear failure of
glued interfaces where failed regions can still transmit load by
remaining in contact.  
\end{abstract}  
\pacs{46.50.+a,62.20.Mk,64.60.-i }
\maketitle
\section{Introduction}
The failure of heterogeneous materials under various types of external
loading conditions has attracted continuous scientific and
technological interest during the past decade
\cite{hh_smfdm,chakrab_beng_book_1997}. Both the macroscopic 
strength and the process of damaging of loaded specimens
strongly depend on the disordered microscopic properties of the
material. Hence, most of the theoretical studies are based on discrete
models which can account for the disordered material properties and
their interaction with the inhomogeneous stress field naturally
arising in a damaged specimen. Fiber bundle models (FBM) are one of
the most important theoretical approaches in this field
\cite{daniels+proc_rsa+1945}, which also  
served as the basis for the development of more complicated
micromechanical models of fracture
\cite{harlow_phoenix_jcm_1978_2,curtin_prl_1998,kun_jmatsc35_2000}. In spite of their 
simplicity, FBMs capture the most important ingredients of the failure
process and make it also possible to obtain several characteristic
quantities of high interest in closed analytic form. Based on FBMs
important results have been obtained for the macroscopic response of
the loaded specimen \cite{pradhan_ijmpb_2003}, and for the temporal \cite{kloster_pre_1997,hansen_crossover_prl,hansen_lower_cutoff_2005,raul_burst_contdam} and spatial structure of damage
on the microlevel. In the framework of FBM the analogy of fracture and
critical phenomena
\cite{andersen_tricrup_prl_1997,moreno_fbm_avalanche,raul_creeprup_twouniv,chakrabarti_phasetrans}
has also been addressed, which is of high 
practical importance for the forecasting of imminent failure of loaded
systems \cite{chakrab_precursor_2002,hansen_crossover_prl}.

Fiber bundle models have also been adopted to study the failure of
glued interfaces of solid blocks \cite{batrouni_intfail_pre_2002,zapperi_crackfuse_eujb_2000,Knudsen_breaksurf_pre_2005,delaplace_jem_2001}. Such interfaces as a
part of complex constructions, are assumed to sustain various types of
external loads. In fiber reinforced composites, where fibers are
embedded in a matrix material, the fabrication of the fiber-matrix
interface strongly affects the mechanical performance of the
composite. Very recently we have shown that under shear loading of 
glued solid blocks, the interface elements may suffer not only
stretching but also bending \cite{frank_ferenc_2005}. In order to capture this effect we
proposed to discretize the interface in terms of beams which can be
elongated and bent, and break due to both deformation modes in a complex
way. During the gradual failure of interfaces of solid blocks under shear,
damaged regions of the interface can still transmit load 
contributing to the overall load bearing capacity of the
interface. This can occur, for instance, when the two solids remain in
contact at the failed regions and exert friction force on each other. 
In many applications the glue between the two interfaces has
disordered properties but its failure characteristics is not perfectly
brittle, the glue under shear may also yield carrying a constant load
above the yield point. 

We present an extension of models of the shear failure of glued
interfaces considering that surface elements after failure
still can have a certain load bearing capacity. The disordered
interface is represented by a parallel set of fibers with random
breaking thresholds and linearly elastic behavior until failure.
The broken fibers are assumed to carry a constant load which is a
fraction $0 \leq \alpha \leq 1$ of their failure load. Varying the
value of $\alpha$ the model interpolates between the perfectly brittle
($\alpha=0$) and perfectly plastic ($\alpha=1$) constitutive behavior of
fibers. Based on analytic calculations and computer simulations, we
show that the finite load bearing capacity of failed fibers has a
substantial effect on both the macroscopic response and microscopic
damage process of the fiber bundle. When the load redistribution
following fiber failure is short ranged, an interesting phase
transition is revealed at a specific value of $\alpha$. 

\section{Model}\label{sec:model}
In order to model the shear failure of glued interfaces, we recently
introduced a model \cite{frank_ferenc_2005} which represents the
interface as an ensemble of 
parallel beams connecting the surfaces of two rigid blocks. The
beams are assumed to have identical
geometrical extensions (length $l$ and width $d$) and linearly elastic
behavior characterized by the Young modulus $E$. In order to capture
the failure of the interface, the beams are assumed to break when
their deformation exceeds a certain threshold value. Under shear
loading of the interface, beams suffer stretching and bending
deformation resulting in two modes of breaking. The 
stretching and bending deformation of beams can be expressed in terms
of a single variable, i.e. longitudinal strain $\varepsilon = \Delta l
/ l$ , which enables us to map the interface 
model to the simpler fiber bundle models. The two breaking modes can be
considered to be independent or combined in the form of a von Mises type
breaking criterion. The strength of beams is characterized by the two threshold
values of stretching $\varepsilon_1$ and bending $\varepsilon_2$ a beam can
withstand. The breaking 
thresholds are assumed to be randomly distributed variables of the
joint probability distribution  
$p(\varepsilon_1, \varepsilon_2)$. The randomness of the breaking
thresholds is supposed to 
represent the disorder of the interface material. After breaking of a
beam the excess 
load has to be redistributed over the remaining intact elements. In
Ref.\ \cite{frank_ferenc_2005} we presented a detailed study which demonstrated 
that the beam model of  
sheared interfaces with two breaking modes can be mapped into a
simple fiber bundle model of a single breaking mode by an appropriate
transformation of the fibers' strength disorder.

\begin{figure}
\psfrag{aa}{{\LARGE $\sigma^i_{th}$}}
\psfrag{bb}{{\LARGE $\alpha\sigma^i_{th}$}}
\psfrag{cc}{{\LARGE $\varepsilon$}}
\psfrag{dd}{{\LARGE $\sigma$}}
  \begin{center}
\epsfig{bbllx=0,bblly=0,bburx=230,bbury=180,file=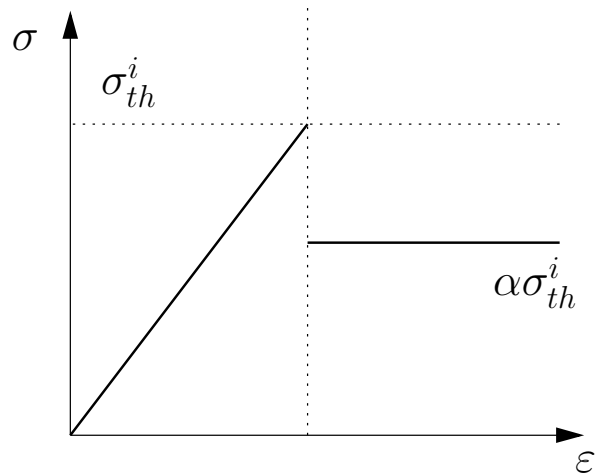,
  width=8.0cm}
   \caption{Constitutive behavior of a single
fiber: the fiber shows linearly elastic behavior up to the breaking
threshold $\sigma^i_{th}$,
then it keeps a fraction $0 \leq \alpha \leq 1$ of the ultimate load
$\alpha\sigma^i_{th}$.
 }   
\label{fig:const_one} 
  \end{center}
\end{figure}

In the present paper, we extend our model by taking into account that failed
surface elements can still carry some external load increasing the load
bearing capacity of the damaged interface. For simplicity, our study
is restricted to discretize the interface in terms of fibers which
could then be further generalized to beams \cite{frank_ferenc_2005}. A bundle
of parallel fibers is considered with breaking thresholds
$\sigma_{th}$ in the interval 
$0\leq \sigma_{th} \leq \sigma_{th}^{max}$ with a probability density
$p(\sigma_{th})$ and distribution function $P(\sigma_{th}) =
\int_0^{\sigma_{th}} p(\sigma_{th}')d\sigma_{th}'$. 
We assume that after the
breaking of a fiber at the failure threshold $\sigma^i_{th}$, it may retain
a fraction $0 \leq \alpha\leq 1$ of
its ultimate load $\sigma^i_{th}$, {\it i.e.} it will continue to
transfer a constant load $\alpha\sigma^i_{th}$ between the surfaces. This
assumption can be interpreted so that at the damaged areas of the
interfaces the two 
solids still remain in contact exerting for instance a friction force
which may contribute to the overall load bearing capacity. 
In many applications the glue between the two interfaces has
disordered properties but its failure characteristics is not perfectly
brittle, the glue under shear may also yield. 
The constitutive behavior of single fibers is
illustrated in Fig.\ \ref{fig:const_one}. Note that the load carried
by the broken fibers 
is independent of the external load, furthermore, it is a random
variable due to the randomness of the breaking thresholds.
Varying the value of $\alpha$, the model interpolates between the
perfectly brittle failure ($\alpha=0$) and perfectly plastic ($\alpha
= 1$) behavior of fibers. The load stored by the failed fibers
reduces the load increment redistributed over the intact fibers, which 
strongly affects the process of gradual failure occurring under
quasi-static loading of the interface. In the following we present a
detailed study of the model system varying the strength of plasticity
$\alpha$. For the range of load sharing the two limiting cases of
global and local load redistributions will be considered after failure
events.

\section{Transition to perfect plasticity}
\label{sec:constit}
Assuming global load sharing (GLS) after fiber breaking,
the constitutive equation of the interface can be cast into a closed
form. At an externally imposed deformation $\varepsilon$ the interface is a
mixture of intact and broken fibers, which both contribute to the
load bearing capacity of the interface. Since the broken fibers retain
a fraction 
$\alpha$ of their failure load, at the instant of fiber breaking only
the reduced load $(1-\alpha)\sigma^i_{th}$ is redistributed over the intact
fibers.
Since the fraction of fibers having breaking threshold in the
interval $[\varepsilon,\varepsilon+d\varepsilon]$ can be obtained as
$p(\varepsilon)d\varepsilon$, the constitutive equation
$\sigma(\varepsilon)$ reads as
\beq{ \label{eq:const}
  \SI(\EP) = \underbrace{\vphantom{\int} E \EP (1- \PPE
)}_{{\displaystyle \SI_{\textrm{DFBM}}}} + \underbrace{\alpha
\int_{0}^{\EP } E \EP ' 
p(\EP ') \,d \EP '}_{{\displaystyle \SI_{\textrm{Pl}}}} , 
  } 
where the integration is performed over the entire load history. The
first term labeled 
$\SI_{\textrm{DFBM}}$ provides the load carried by the intact fibers,
which corresponds to the classical dry fiber bundle (DFBM) behavior
\cite{daniels+proc_rsa+1945,kloster_pre_1997,sornette_jpa_1989}.  
\begin{figure}

\epsfig{bbllx=0,bblly=110,bburx=580,bbury=760,file=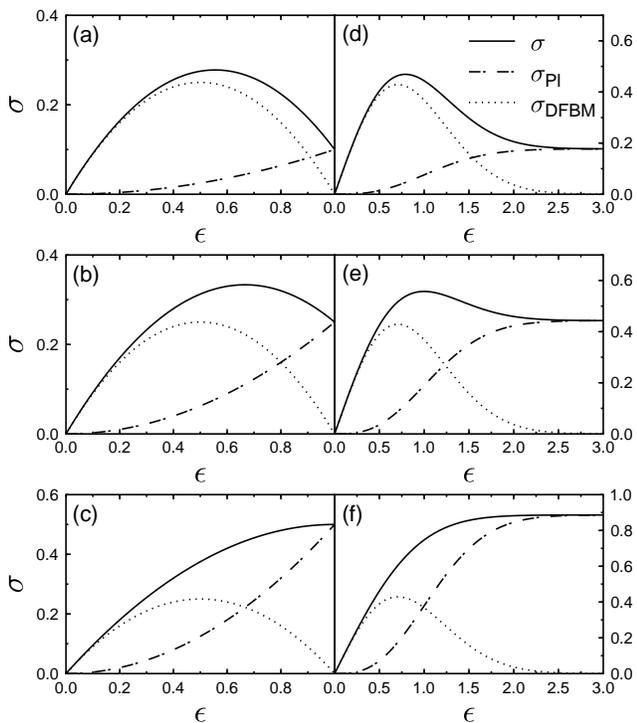,
  width=8.7cm}
\caption{Constitutive behavior $\SI (\EP )$ of the plastic fiber
bundle for uniform ($a$, $b$, $c$) and Weibull distribution with $m=2$
($d$, $e$, $f$) at $\AL=0.2$ ($a$, $d$),  $\AL=0.5$ ($b$, $e$) and $\AL=1.0$ ($c$, $f$).
The contribution of intact $\SI_{\textrm{DFBM}}$ and
failed fibers $\SI_{\textrm{Pl}}$ are also shown. Note that
$\SI_{\textrm{DFBM}}$ is identical with the constitutive curve of
simple dry fiber bundle models. 
}
\label{fig:constit}
\end{figure}
The constitutive law of DFBMs is recovered in the limiting case $\AL
=0$, when the complete load of the failed fiber is transferred
to the remaining intact fibers of the bundle. 
In the second term $\sigma_{Pl}$, which accounts for the load carried
by the broken fibers, the integral is calculated over the entire load
history of the interface 
up to the macroscopic deformation $\varepsilon$. It can be seen in
Eq.\ (\ref{eq:const}) that the value of $\alpha$ controls the relative importance
of the {\it elastic} and {\it plastic} terms influencing the
macroscopic response $\sigma(\varepsilon)$ and also the microscopic
damage process of the system. When  
$\alpha$ is increased, less load is transfered to the intact fiber and
in the limiting case $\alpha=1$ 
failed fibers retain their entire load so no load transfer
occurs. In this report, we explore the influence of the parameter
$\AL$ when it is tuned between these two extremal cases. In the
following calculations the value of the fibers' Young modulus was set
to unity $E=1$. 

We note that the plastic fiber bundle model resembles up to some extent
to the continuous damage fiber bundle model (CDFBM) worked out in
Refs.\ \cite{kun_epjb_2000,raul_burst_contdam}. The main assumption of
the CDFBM is that due to the activation of certain internal degrees of
freedom, the fibers undergo a gradual softening process reducing their
Young modulus in consecutive partial failure events. The fibers always
remain linearly elastic but with a 
Young modulus $E(k) = a^kE$, where the multiplication factor $0 \leq
\alpha \leq 1$ describes the stiffness reduction in a single failure
event and $k$ denotes the number of failures occurred. If the fibers
can fail only once ($k=1$) and keep their stiffness value constant,
the constitutive law of the system reads 
as  
\beq{\label{eq:cdam}
\SI (\EP) = E\EP (1- \PPE ) + aE \EP \PPE.
}
It was demonstrated in Refs.\ \cite{kun_epjb_2000,raul_burst_contdam}
that increasing the number of times $k$ the fibers can fail, the CDFBM develops
a plastic plateau, however, with a mechanism completely different from
the one considered here.

\begin{figure}
  \begin{center}
\epsfig{bbllx=110,bblly=350,bburx=430,bbury=640,file=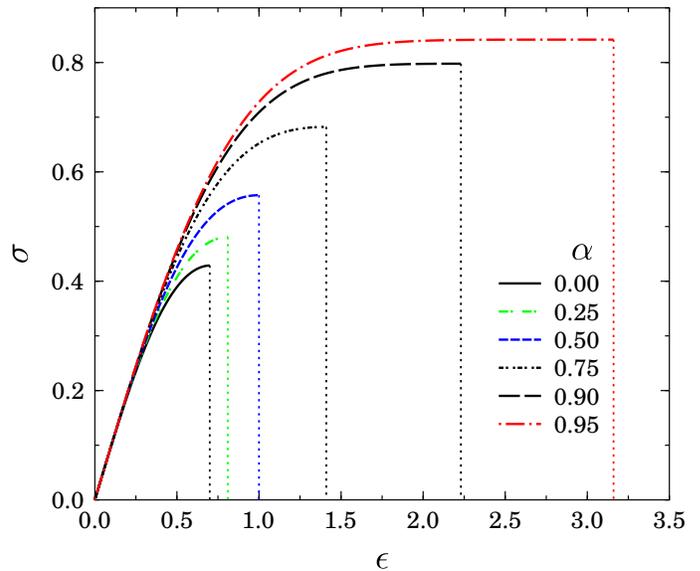,
  width=8.8cm}
    \caption{{\it (Color online)} Simulations of stress controlled loading of a bundle of
$N=1.6\times 10^6$ fibers with Weibull distributed breaking thresholds
($\lambda=1, m=2$). For clarity, the occurrence of macroscopic failure
is indicated by vertical lines. Increasing $\alpha$ the constitutive behavior
becomes perfectly plastic.}
\label{gr:const2} 
  \end{center}
\end{figure}
It is instructive to consider two fundamentally different cases of
disorder distributions $\PPE$, namely bounded and unbounded ones,
where the largest breaking threshold $\sigma_{th}^{max}$ takes a
finite value or goes to infinity, respectively. In this report, we focus
on two specific realizations, {\it i.e.} a uniform distribution
between 0 and $\sigma_{th}^{max}$ 
\beq{ \label{eq:uniform}
  P(\sigma_{th}) = \frac{\sigma_{th}}{\sigma_{th}^{max}} , \ \ \ \ \ 0
\leq \sigma_{th} \leq \sigma_{th}^{max},  
  }
and distributions of the Weibull type
 \beq{ \label{eq:weibull}
   P(\sigma_{th})  = 1- e^{- \left( \sigma_{th} / \lambda \right)^m} ,
   }
are considered where $\lambda$ and $m$ denote the characteristics
strength and Weibull modulus of the distribution, respectively. 
For our study the Weibull distribution has the advantage that the
amount of disorder in the failure thresholds can easily be controlled
by the value of $m$.

The functional form of the constitutive behavior $\sigma(\EP)$ is
shown in Fig.\ \ref{fig:constit} for both disorder
distributions Eqs.\ (\ref{eq:uniform},\ref{eq:weibull}). 
It is interesting to note that for $\alpha <1$ there always exists a
maximum of $\SI(\EP )$, just as in the case of DFBM. Under stress
controlled loading conditions, macroscopic failure occurs at the
maximum of $\sigma(\EP)$ so that the position and  
value of the maximum define the critical stress $\SC$ and strain
$\EC$ of the bundle, respectively. It can be observed in Fig.\
\ref{fig:constit} 
that the value of $\SC$ and $\EC$ are both higher than the
corresponding values of DFBM indicating that the presence of plastic
fibers increases the macroscopic strength of the bundle.
The decreasing part and the plateau of $\sigma(\varepsilon)$ can be
realized under strain 
controlled loading conditions gradually increasing $\EP$. Under strain
control the local load on the fibers is 
determined by the externally imposed deformation so that there is no
load redistribution after fiber failure. The fibers break one-by-one
in the increasing order of their failure thresholds
$\SI_{th}^i=E\varepsilon_{th}^i$. When 
the deformation $\EP$ approaches the maximum value of the breaking
thresholds $\EP_{th}^{max} = \SI_{th}^{max}/E$,
all fibers must fail gradually so that the load of intact fibers
$\SI_{\rm DFBM}$ tends to zero, while that of the broken fibers
$\SI_{\rm Pl}$ takes a finite asymptotic value
\beq{ \label{eq:asymp_sig}
\SI_{\rm Pl} \rightarrow \tilde{\sigma} = \alpha E
\int_0^{\infty}\EP'p(\EP')d\EP' = 
\alpha\left<\SI_{th}\right>,
}
where the integral is equal to the average fiber strength $\left<\SI_{th}\right>$.   
\begin{figure}
  \begin{center}
\epsfig{file=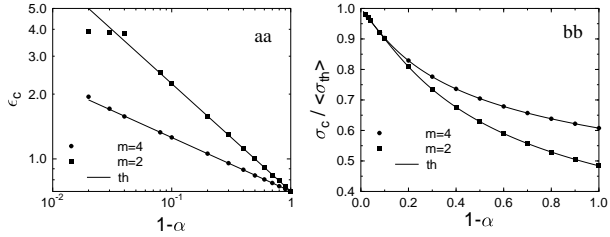, width=\linewidth}
    \caption{Critical strain $\EC$ $(a)$, and critical stress $\SC$ $(b)$
as a function of $1-\AL$ for a Weibull distribution 
with $\lambda=1, m=2$; symbols: simulation results, solid lines:
analytical expressions Eqs.\
(\ref{eq:eps_crit},\ref{eq:converg},\ref{eq:aver_strength}). 
}
  \label{gr:escrit}
  \end{center}
\end{figure}
When the strength of plasticity $\AL$ is increased, the critical
strain $\EC$ and stress $\SC$, furthermore, the asymptotic stress of
the plateau $\tilde{\sigma}$ increase. 
The value of the critical deformation $\EC$ can be obtained by
differentiating Eq.\ (\ref{eq:const}) with respect to $\EP$ and
calculating the root \cite{sornette_jpa_1989}
\beq{\label{eq:root}
1-P(\EC)-\EC p(\EC)\left[1-\alpha\right] = 0,
} 
from which the critical stress follows as $\sigma_c = \sigma(\EC)$.
Eq.\ (\ref{eq:root}) implies that in the limiting case of
$\alpha \rightarrow 1$ the critical strain $\EC$ tends to the maximum of the
breaking thresholds $\EP_{th}^{max}$, where $P(\EP_{th}^{max})=1$. For the
uniform distribution Eq.\ (\ref{eq:uniform}) we obtain
\beq{ \label{eq:ecrit_w12}
  \EC = \frac{\EC^{0}}{1-\alpha/2},  \ \ \ \ \mbox{hence,} \ \ \ \ \EC
\xrightarrow[\alpha \rightarrow 1]{} 2\EC^{0}=\EP_{th}^{max}.
}
Here $\varepsilon_c^0$ denotes the critical strain of DFBM
$\varepsilon^0_c = \varepsilon_{th}^{max}/2$, which can be obtained by
setting $\alpha=0$ in Eq.\ (\ref{eq:root}).
It follows that for unbounded threshold distributions like the Weibull
distribution, $\EC$ diverges so that perfect plasticity is only
reached in the limit $\EC \rightarrow \infty$. 
The functional form of the divergence is
not universal, due to the structure of the third term on the left hand
side of Eq.\ (\ref{eq:root}), $\EC$ depends on the specific form of $p(\EP)$.
For the Weibull distribution, $\EC$ as a function of $\alpha$ reads as 
\beq{
  \EC = \EC^{0}\left(1-\AL\right)^{-1/m}, \ \ \ \mbox{where} \ \ \
\EC^{0} = \lambda\left(\frac{1}{m}\right)^{1/m} 
\label{eq:eps_crit}
  }
for any Weibull exponent $m$. 
Parallel to this, the decreasing part and the plateau of the
constitutive curve $\sigma(\EP)$ disappear so that $\SC$ and
$\tilde{\sigma}$ converge to the same finite value, which is the
average fiber strength $\left<\SI_{th}\right>$
\beqa{\label{eq:converg}
\tilde{\sigma} \rightarrow \left<\SI_{th}\right> \ \ \ \ \ \mbox{and}
\ \ \ \ \ \SC \rightarrow \left<\SI_{th}\right>. 
}
The average fiber strength $\left<\SI_{th}\right>$ can be determined as 
\beqa{\label{eq:aver_strength}
\left<\SI_{th}\right> = \frac{\SI_{th}^{max}}{2} \ \ \ \ \ \
\mbox{and} \ \ \ \ \ 
\left<\SI_{th}\right> = \frac{1}{m}\Gamma\left( \frac{1}{m}\right)
}
for the uniform and Weibull distributions, respectively. Here $\Gamma$
denotes the Gamma function.

In order to illustrate this behavior,  
Fig.\ \ref{gr:const2} presents constitutive curves for Weibull
distributed fiber strength obtained by computer simulations of stress
controlled loading up to the critical point with $\lambda=1$ and
$m=2$. It is apparent that in the
limiting case of $\alpha \rightarrow 
1$ the constitutive curve $\SI(\EP)$ reaches a plateau,
indicating a perfectly plastic macroscopic state of the system.
The position of the maximum $\EC$ of the constitutive curves, {\it
i.e.} the ending point of the curves, rapidly increases as $\alpha$
approaches 1, while the value of the maximum $\SC$ tends to a finite
value. In agreement with the analytic predictions Eq.\
(\ref{eq:eps_crit}), simulations confirmed that $\EC$ diverges as a power
law whose exponent depends on the parameters of the strength
distribution (see Fig.\ \ref{gr:escrit}). 

\begin{figure}
  \begin{center}
\epsfig{bbllx=90,bblly=350,bburx=440,bbury=630,file=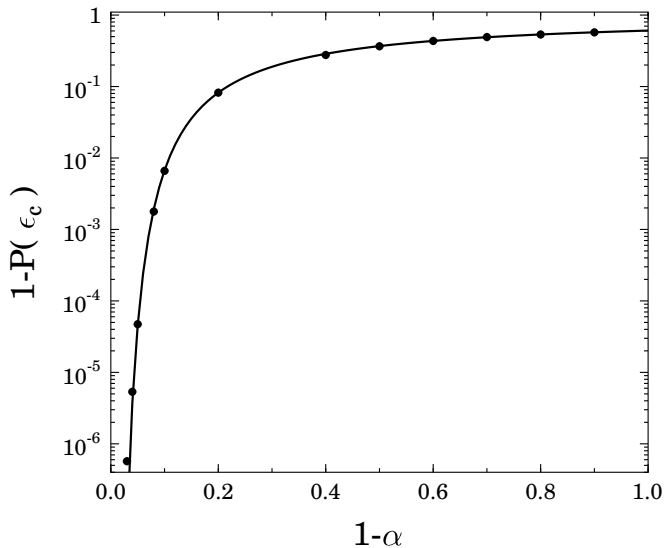,
  width=9.0cm}
    \caption{Fraction of intact fibers $\phi=1-P(\EC(\alpha))$ at the point of
macroscopic failure $\EC$ vs. $1-\AL$ for a Weibull distribution of
$\lambda=1, m=2$; 
circles: GLS simulation results, solid line: analytical solution Eq.\
(\ref{eq:phi_alpha}).} 
\label{gr:pec} 
  \end{center}
\end{figure}

Controlling the external stress, the constitutive curve of the system
Fig.\ \ref{gr:const2} can only be realized up to the maximum, since at the
critical load $\SC$ abrupt failure of the bundle occurs breaking all
the surviving intact fibers in a large burst. The fraction $\phi$ of
fibers which break in the final burst causing global failure can be
determined as $\phi = 1-P(\EP_c(\alpha))$, which is illustrated in
Fig.\ \ref{gr:pec} as a function of $1-\alpha$ for the specific case
of a Weibull distribution 
\beq{\label{eq:phi_alpha}
  \phi(\alpha) = e^{-1/m(1- \AL)}.
  }
It can be observed that as the system approaches the state of perfect
plasticity $\alpha \rightarrow 1$, $\phi$ tends to zero. This
demonstrates that more and more fibers break before global failure
occurs, and perfect plasticity is obtained when the strongest fiber
fails at the maximum of $\SI(\EP)$ (compare also to Fig.\
\ref{gr:const2}). This argument 
also implies that for $\alpha \rightarrow 1$, the difference of the
microscopic damage process under stress and strain controlled loading
disappears, the fibers break one-by-one without triggering avalanches
of breakings.

\section{Avalanches of fiber breakings}
Under stress controlled loading of the fiber bundle, the load dropped
by a breaking fiber is redistributed over the intact ones. This load
increment can give rise to further breakings which then may trigger an
entire avalanche of failure events. 
The distribution $D( \Delta)$ of avalanche sizes $\Delta$ is an
important quantity for the dynamical description of the loaded system.
For the case of classical DFBMs ($\alpha = 0$) under GLS conditions the
avalanche size 
distribution $D(\Delta)$ can be obtained analytically
\cite{hemmer_distburst_jam_1992,kloster_pre_1997} as an
integral, from which the asymptotic form of the distribution for large
avalanches proved to be a power law  
\beq{ \label{eq:delta5_2}
  D(\Delta) \propto \Delta^{-5/2}, \, \ \ \ \ \ \ \Delta \rightarrow \infty .
}
\begin{figure}
  \begin{center}
\epsfig{bbllx=90,bblly=340,bburx=440,bbury=640,file=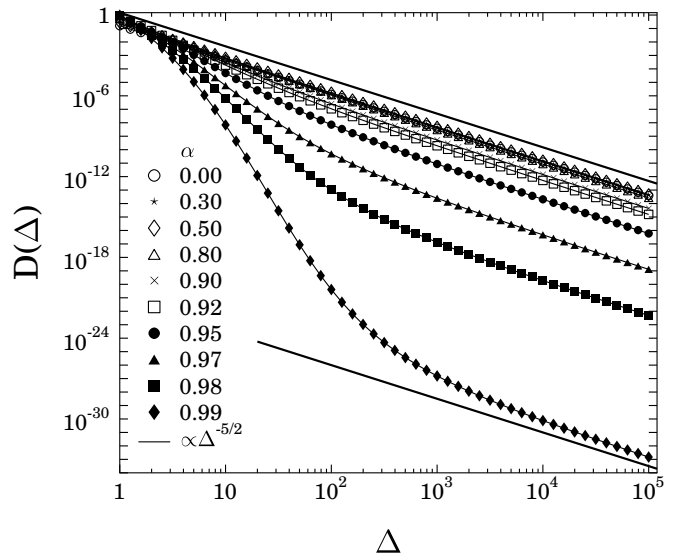,
  width=9.0cm}
\caption{ Analytic solution of the avalanche size distribution
$D(\Delta)$ at various different values of $\alpha$. For $\alpha \to
0$ the usual power law distribution is recovered, whereas for $\alpha \to
1$ an exponential decay of $D(\Delta)$ is obtained.    
For the specific calculations a Weibull distribution was used with
$m=2$. }  
\label{gr:gls_avals:a} 
\end{center}
\end{figure}
The value of the exponent $5/2$ is universal, it does not depend on
the details of the disorder distribution of the failure thresholds
\cite{hemmer_distburst_jam_1992,kloster_pre_1997}. 

\begin{figure}
  \begin{center}
\epsfig{bbllx=90,bblly=350,bburx=440,bbury=630,file=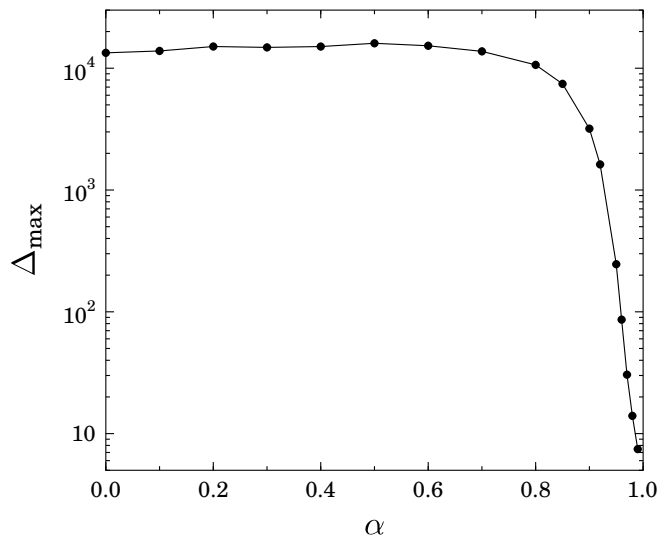,
  width=9.0cm}
    \caption{Size of the largest avalanche in a GLS simulation with a
Weibull distribution of $m=2$. }  
\label{gr:largest_resc} 
  \end{center}
\end{figure}

In order to obtain the analytical solution for the avalanche
distribution in the presence of plastic fibers  $\alpha \neq
0$, we can follow the derivation of Refs.\
\cite{hemmer_distburst_jam_1992,kloster_pre_1997}, taking into account
that the average number of fibers $a(\EP,\AL)d\EP$ which break as a
consequence of the load increment caused by a fiber breaking at the
deformation $\EP$, is reduced by a factor of $(1-\alpha)$ 
\beq{
  a(\EP, \AL)d\EP = \frac{\EP p(\EP) (1- \AL) }{1-P(\EP)}d\EP .
}
Taking into account that the critical deformation $\EC$ where macroscopic
failure occurs also depends on $\alpha$, the avalanche size
distribution $D(\Delta)$ can be cast in the form
\beqa{\label{eq:aval_alpha_gen}
  \frac{D(\Delta)}{N} &=& \frac{\Delta^{\Delta-1}}{\Delta !} \int
\limits_{0}^{\EC(\AL)} a(\EP, \AL)^{\Delta -1} e^{-a(\EP, \AL)
\Delta} \times \\ [2mm] \nonumber
&&[1-a(\EP,\AL)]p(\EP) \,d\EP . 
}
For the specific case of the Weibull distribution
with an arbitrary modulus $m$ the general equation Eq.\ (\ref{eq:aval_alpha_gen})
can be written in the form 
\begin{eqnarray}
\label{eq:aval_alpha}
  D(\Delta, \AL) &=& \frac{\Delta^{\Delta-1}   }{ \Delta ! (m(1-\AL
)) ^2 \Delta _c ^{\Delta +1}}\left[ \gamma ( \Delta, \Delta _c)\right. \\ [2mm]
&+& \left. \Delta _c^{\Delta} m (1- \AL ) e^{- \Delta _c} \right] , \nonumber
\end{eqnarray}
where $\Delta_c$ depends on the amount of disorder $m$ and on the
strength of plasticity $\alpha$
\begin{equation}
\label{eq:aval_alpha_d}
 \Delta _c = \Delta + \frac{1}{ m(1- \AL)} .
\end{equation}
In Eq.\ (\ref{eq:aval_alpha}) $\gamma$ denotes the incomplete Gamma
function \footnote{There are several definitions of the incomplete
Gamma function, we use $ \gamma (a,x) = 
\int \limits_{0}^{x} e^{-t}t^{a-1} \,dt$ .}. 
Two limiting cases can be distinguished in the solution: first, for 
$\AL \rightarrow 0$ the classical power law dependence 
Eq.\ (\ref{eq:delta5_2}) is recovered. This analytic solution is
illustrated in Fig.\ \ref{gr:gls_avals:a} for a Weibull distribution
with $m=2$, where a power law of $D(\Delta)$ is apparent for $\alpha <
0.9$.  
\begin{figure}
  \begin{center}
\epsfig{bbllx=90,bblly=340,bburx=440,bbury=640,file=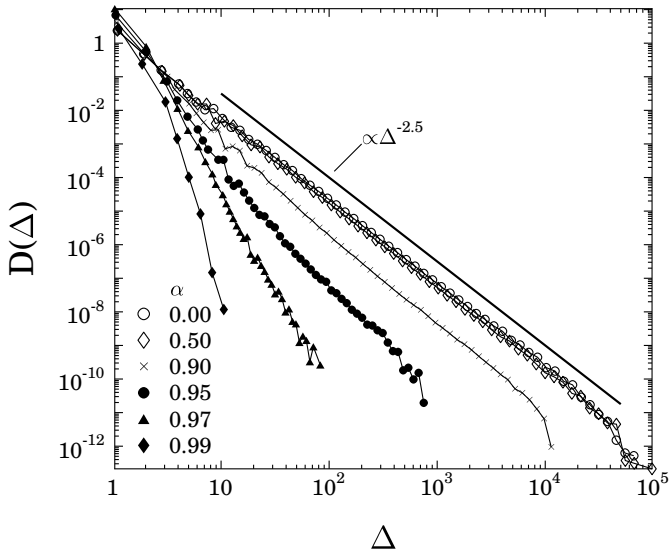,
  width=9.0cm}
    \caption{ Distribution $D(\Delta)$ of avalanches of size $\Delta$
for various values of $\AL$ obtained by computer simulations for a
system of $N=1.6\cdot 10^7$ fibers with Weibull distributed failure
thresholds $m=2$. Satisfactory agreement is obtained with the
analytic results presented in Fig.\ \ref{gr:gls_avals:a}.
} 
\label{gr:gls_avals:b} 
  \end{center}
\end{figure}
However, for the limiting case of $\AL \rightarrow 1$, we have to
consider the behavior of the argument $\Delta _c$ of the  analytic
solution Eq.~(\ref{eq:aval_alpha}). For $\AL \approx 1$, there
will be a regime of $\Delta$ values where the term $1/(m (1- \AL))$
dominates over $\Delta$ resulting in a faster decay of the
distribution $D(\Delta)$ than any power. Still, for any values of
$\alpha$ in the limiting case $\Delta >> \Delta_c(\alpha)$,
the usual mean field power law behavior Eq.\
(\ref{eq:delta5_2}) is asymptotically recovered.
Avalanche size distributions $D(\Delta)$ obtained from computer
simulations at various different values of $\alpha$ are presented in
Fig.\ \ref{gr:gls_avals:b}. 
In a good quantitative agreement with the analytic predictions, the
numerical results can be well fitted by a power law of exponent 5/2
for moderate values of $\alpha$.
However, for $\AL >0.9$ strong deviations from the power law
Eq.~(\ref{eq:delta5_2}) can be observed for intermediate
avalanche sizes $1 \leq \Delta \leq 10^3$, which appears to be
an exponential decay. 
Although in the analytical solution the asymptotic power law
behavior is still visible for very large $\Delta$, see
Fig.~\ref{gr:gls_avals:a}, computer simulations in Fig.\
\ref{gr:gls_avals:b} show solely a very steep decrease. 
It can be seen in the analytic solution in Fig.\ \ref{gr:gls_avals:a}
that the relative frequency of avalanches 
of size $\Delta > \mathcal{O}(10^3)$ is $D =
\mathcal{O}(10^{-30})$ for $\alpha=0.99$, so it would require
extremely large systems to count any such events.   
The size of the largest avalanche $\Delta_{max}$ is plotted in Fig.\
\ref{gr:largest_resc} as a function of $\alpha$. Obviously,
$\Delta_{max}$ is a monotonically decreasing function of $\alpha$
whose decrease gets faster in the regime where the distribution
$D(\Delta)$ exhibits the crossover to the faster decaying form.

An important consequence of the analytic solution Eqs.\
(\ref{eq:aval_alpha},\ref{eq:aval_alpha_d}) is that the characteristic
avalanche size where the crossover occurs from a power law to a faster
decaying exponential form also depends on the amount of disorder, {\it
i.e.} the stronger the disorder is, the larger
the crossover size gets at a given $\alpha$.

\section{Local load sharing}
From experimental and theoretical point of view, it is very important
to study the behavior of the plastic bundle when the interaction of
fibers is localized. 
In the case of local load sharing (LLS) under stress controlled
external loading conditions, the load
dropped by the broken fiber is redistributed in a local neighborhood of
the fiber giving rise to high stress concentration in the vicinity of
failed regions. Stress concentration leads to correlated growth of
clusters of broken fibers (cracks), which plays a
crucial role in the final 
breakdown of the system, {\it i.e.} macroscopic failure of the bundle
occurs due to the instability of a broken cluster which then triggers
an avalanche of failure events where all the remaining intact fibers
break. This effect typically leads to a more brittle constitutive
behavior of the system and the appearance of non-trivial spatial and
temporal correlations in the damage process
\cite{hansen_distburst_local_1994,batrouni_intfail_pre_2002,raul_varint_2002,kun_epjb_2000}.

In the plastic bundle, after a fiber breaks it still retains a fraction
$\AL$ of its failure load $\SI_{th}$ so that only the amount
$(1-\AL)\SI_{th}$ is redistributed over the intact fibers in the
neighborhood. It implies that the load bearing broken fibers reduce
the stress concentration around failed regions giving rise to
stabilization which also affects the temporal and
spatial evolution of damage during the loading process.

\begin{figure}
  \begin{center}
\epsfig{file=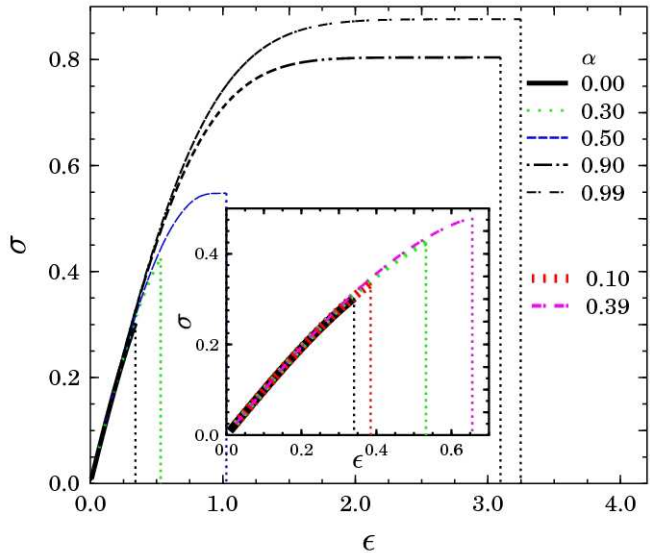,  width=8.7cm}
   \caption{{\it (Color online)} Constitutive law  $\SI(\EP)$ of the LLS bundle obtained by
computer simulations of a system of size $L=401$ for several different
values of $\AL$. The inset  
shows a magnified view of $\SI(\EP)$ for the regime $\alpha <
0.4$. For clarity, vertical lines indicate the location of macroscopic
failure. For the breaking thresholds a Weibull distribution was used with
$m=2$.}  
\label{gr:const_lls1} 
  \end{center}
\end{figure}
In the following we consider a bundle of $N$ fibers organized on a
square lattice of size $L\times L$ with periodic boundary
conditions. The fibers are assumed to have Weibull distributed strength
Eq.\ (\ref{eq:weibull}), where the value of $\lambda$ is always set to
unity and for the Weibull modulus two different values are considered:
$m=2$ (large disorder) and $m=4$ (smaller disorder). After a failure event the
load dropped by the broken fiber $(1-\alpha)\SI_{th}^i$ is equally 
redistributed over the nearest and next-nearest intact neighbors
in the square lattice, {\it i.e.} the local neighborhood of
a broken fiber contains at most 8 intact sites.  
Stress controlled simulations have been carried out for system sizes
ranging from $L=33$ to $L=801$ varying the strength of plasticity
$0\leq \alpha \leq 1$. 

\subsection{Macroscopic response}
It has been shown for DFBMs where broken fibers carry no load, that the
macroscopic response of the bundle 
when the interaction of fibers is localized follows the
constitutive law of the corresponding GLS system with a reduced
critical strain and stress, {\it i.e.} 
the $LLS$ bundle behaves macroscopically in a more brittle way than
its GLS counterpart
\cite{hansen_distburst_local_1994,kun_epjb_2000,raul_varint_2002}. 
Figure \ref{gr:const_lls1} shows the constitutive curve of a plastic bundle of
size $L=401$ for several different values of $\alpha$.
It can be observed that for $\alpha \approx 0$ the constitutive curve
exhibits the usual $LLS$ behavior, {\it i.e.} the macroscopic failure
is preceded by a relatively short non-linear regime and global
failure occurs in an abrupt manner. The position of the macroscopic
failure defines the value of the critical strain $\EC^{LLS}$ and stress
$\SC^{LLS}$. It is very interesting to note that when $\alpha$ is
increased, the LLS constitutive curves practically recover the
behaviour of the corresponding GLS system, {\it i.e.} for $\alpha \geq 0.4$
the macroscopic failure occurs when reaching the plateau of
$\SI(\EP)$. 

\begin{figure}   
  \begin{center}
\epsfig{bbllx=160,bblly=320,bburx=500,bbury=610,file=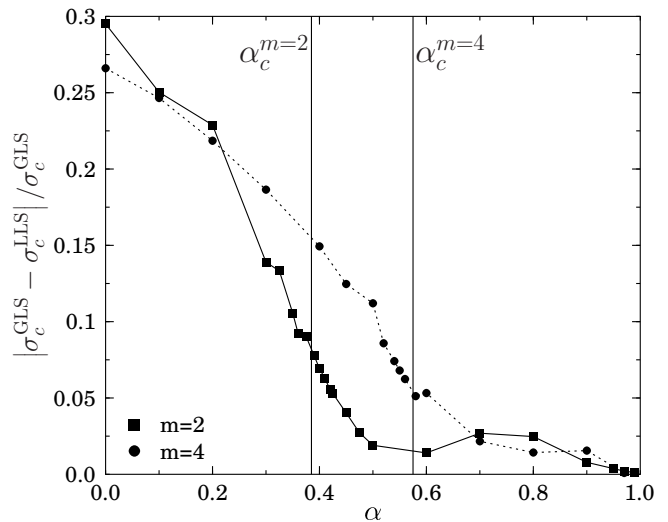,
  width=8.7cm}
\caption{
The relative difference of the critical stresses
$\sigma_c^{GLS}$ and $\sigma_c^{LLS}$ of global and local load sharing
systems as a function of $\alpha$ for two
different values of the Weibull modulus $m$. The vertical lines
indicate the critical values of $\alpha$, which were obtained in Sec.\
\ref{sec:spatial}.}
\label{gr:scdiff}
\end{center}
\end{figure}
The convergence of the LLS system to the GLS macroscopic behavior is
better seen in Fig.\ \ref{gr:scdiff} where the relative difference of the
critical stresses $\SC^{GLS}(\alpha)$ and $\SC^{LLS}(\alpha)$ of the
global and local load sharing bundles is presented. 
It can be seen
in the figure that there exists a threshold value $\alpha_c$ of $\alpha$ above
which the macroscopic response of the LLS bundle becomes very close to
the corresponding GLS system, while below $\alpha_c$ the constitutive
behavior of the bundle changes continuously from the usual LLS
response with a high degree of brittleness ($\alpha=0$) to the global load
sharing behavior. It seems that at $\alpha_c$ a continuous transition occurs between the
two regimes. The transition indicates that as a consequence of the
reduction of stress concentration around failed fibers, the bundle can
sustain higher external loads and is able to keep its integrity until
the maximum of $\sigma(\EP)$ is reached.

\begin{figure}
  \begin{center}
\epsfig{bbllx=90,bblly=340,bburx=440,bbury=640,file=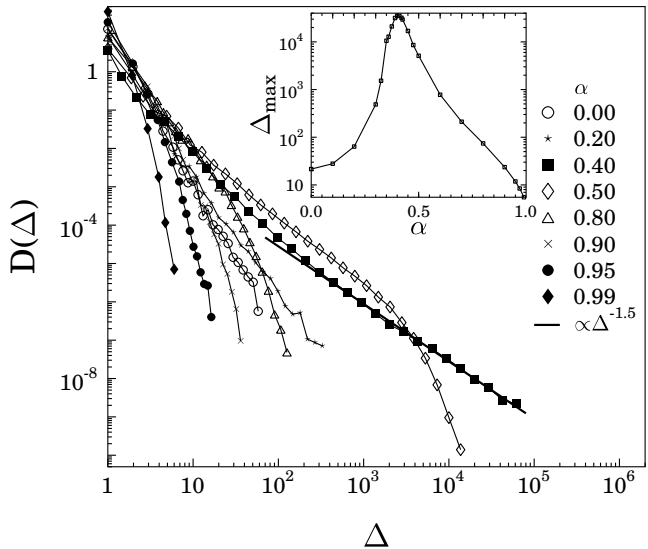,
  width=9.0cm}
   \caption{Avalanche size distributions $D(\Delta)$ obtained by
computer simulations for the system size $L=401$ with local load
sharing, using Weibull distributed failure thresholds $m=2$. The power
law fit is demonstrated for $\alpha =0.4$. In the
inset the largest avalanche $\Delta_{max}$ is plotted versus $\alpha$.}
\label{gr:lls_avals}  
  \end{center}
\end{figure}
\subsection{Bursts of fiber breakings}
The evolution of the macroscopic response of the system with
increasing $\alpha$ is accompanied by
interesting changes of the damage process on the micro-level,
characterized by the avalanches of fiber breakings and the cluster
structure of failed regions. 
The avalanche statistics presented in Fig.\ \ref{gr:lls_avals} shows
remarkable features.
For $\alpha \approx 0$, due to the high stress concentration around
failed fibers, the LLS bundle can only tolerate small avalanches so that
the avalanche size distribution $D(\Delta)$ decays rapidly. With
increasing $\AL$ the higher amount of load kept by broken fibers can
stabilize the bundle even after larger bursts,
hence, the cut-off of the distributions moves to higher
values. It is interesting to note that also the functional form of the
distribution  $D(\Delta)$ changes, {\it i.e.} when $\AL$ approaches
$\AL_c$ the exponential cut-off disappears and the distribution
becomes a power law 
\beq{
D(\Delta) \sim \Delta^{-\mu}
}
for large avalanches. The exponent $\mu$ of the power law was determined
numerically as $\mu^{LLS}=1.5\pm 0.07$, which is significantly lower than the
mean field value $\mu^{GLS} = 2.5$ \cite{kloster_pre_1997}.
Increasing $\alpha$ above the critical point an exponential cut-off
occurs and the power law regime of large avalanches gradually
disappears. Comparing Fig.\ \ref{gr:lls_avals} to the corresponding
GLS results presented in Fig.\ \ref{gr:gls_avals:b}, it is apparent
that above $\alpha_c$ the LLS distributions $D(\Delta)$ have the same
functional form and follow the same tendency with increasing $\alpha$
as the mean field results. It can be concluded that the avalanche
statistics presents the same transitional behavior between the local
load sharing and mean field regimes as observed for the macroscopic
response. The same value of $\mu^{LLS}$ was obtained numerically for
$m=4$, indicating the universality of the exponent with
respect to the strength of disorder. 
The transition is more evident in the inset of Fig.\
\ref{gr:lls_avals}, where the size of the largest avalanche
$\Delta_{max}$ is plotted as a function of $\AL$. The sharp peak
indicates the transition point whose position defines 
$\alpha_c$, while in GLS the largest avalanche $\Delta_{max}$ was a
monotonically decreasing smooth function (compare to Fig.\
\ref{gr:largest_resc}). 
\begin{figure}
\psfrag{aa}{{\large $a)$}}
\psfrag{bb}{{\large $b)$}}
\psfrag{cc}{{\large $c)$}}
\psfrag{dd}{{\large $d)$}}
  \begin{center} 
\epsfig{bbllx=0,bblly=0,bburx=545,bbury=535,file=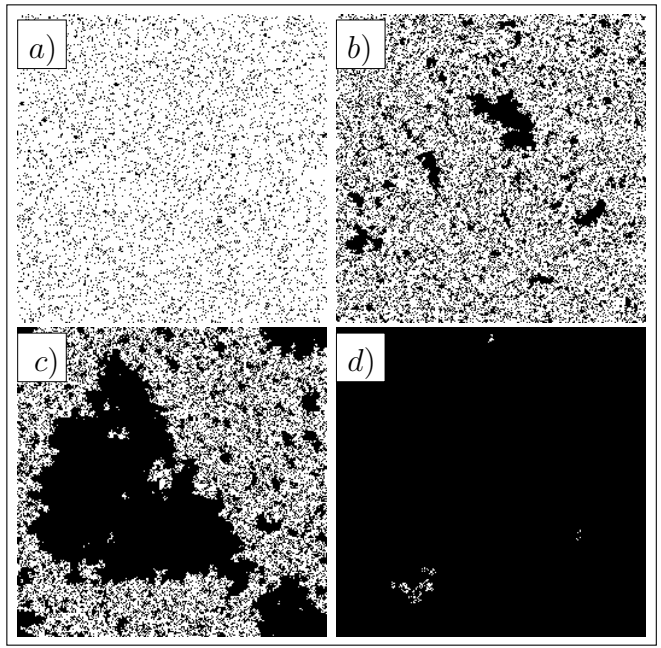,
  width=8.7cm}
    \caption{Latest stable configuration in LLS simulations of
a system of size $L=401$, with a Weibull strength distribution $m=2$
at different values of the control parameter $\alpha$
(a) $0.0$, (b) $0.35$, (c) $0.4$, (d)
$0.6$. The strength of the largest cluster $P_{\infty}$ in the
lattices are (a) $0.003$, (b) $0.097$,
(c) $0.517$, (d) $0.999$. Broken and intact fibers are indicated by
black and white, respectively.
}\label{gr:pattern} 
\end{center}    
\end{figure}

\subsection{Spatial structure of damage}
\label{sec:spatial}
Gradually increasing the external load in the fiber bundle, the
weakest fibers break first in an uncorrelated manner. Since the load is
redistributed solely over the intact neighbors of the broken fiber, the
chance of fiber breakings increases in the vicinity of damage
regions. This effect can result in correlated growth of clusters of
broken fibers with a high stress concentration around their
boundaries. The larger the cluster is, the higher stress concentration
arises. Global failure of the bundle occurs when, due to an
external load increment, one of the clusters becomes unstable and grows until all fibers break. The spatial structure of the damage
emerging when the interaction of fibers is localized can be
characterized by studying the statistics and structure of clusters of
broken fibers. Former studies of the limiting case of very localized
interactions have revealed that the size of the largest cluster in the
system is rather limited, furthermore, it is independent of the system
size. Since the clusters are relatively small, merging of neighboring
clusters does not occur frequently. The clusters themselves are found
to be compact objects 
dispersed homogeneously over the cross section of the bundle
\cite{zapperi_physa270_1989,zapperi_pre59_1999,kun_epjb_2000}. 

\begin{figure}
  \begin{center}
   \epsfig{bbllx=90,bblly=340,bburx=440,bbury=640,file=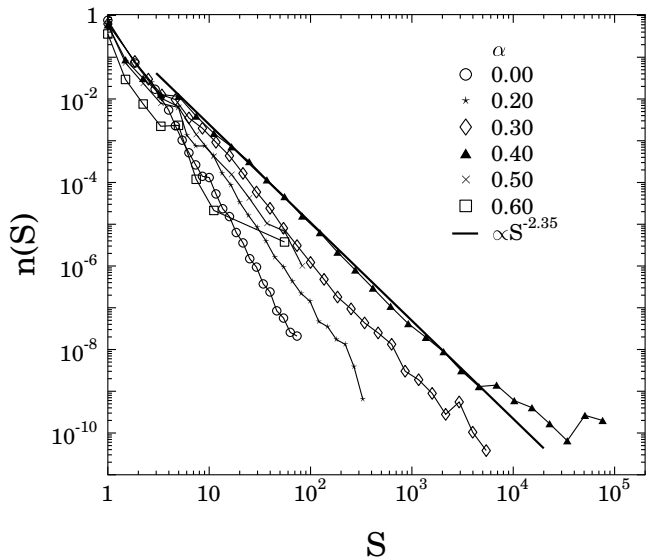,
  width=9.0cm}
   \caption{Distribution $n$ of the size $S$ of broken clusters in
LLS simulations with a Weibull distribution $m=2$, for
different values of $\AL$. The spanning clusters were excluded from the
distributions for $\alpha \geq 0.4$.
}   
\label{gr:clustsizedist}
  \end{center}
\end{figure}
In Fig.~\ref{gr:pattern} the latest stable configuration of the bundle
is presented just before catastrophic failure occurs at the critical
load $\sigma_c^{LLS}$ for several different values of $\AL$.
For $\AL \approx 0$ we note only small
clusters of broken fibers as it is expected for LLS bundles (Fig.\
\ref{gr:pattern}a). With increasing $\AL$, these clusters grow and
adjacent clusters can even merge further increasing the typical
cluster size (Fig.\ \ref{gr:pattern}b). Around the
critical value of $\AL \approx 0.4$, a {\it spanning cluster} of broken fibers
seems to appear (Fig.\ \ref{gr:pattern}c), whereas for higher values of $\AL
>0.4$ almost all fibers have failed (Fig.\ \ref{gr:pattern}d) already
by the time the critical stress is reached.   
The existence of very large clusters is the direct consequence of the
increased load bearing capacity of broken fibers.

Clusters of broken fibers were identified in the square lattice using
the Hoshen-Kopelman algorithm. We evaluated the distribution of cluster
sizes $n(S)$ in the last stable configuration just before macroscopic
failure occurs.
The behavior of $n(S)$ shows again the transitional nature we have
observed for other quantities. It can be seen again in Fig.\
\ref{gr:clustsizedist} that a well defined $\alpha_c$ exists which
separates two regimes: for $\alpha < \alpha_c$ the clusters are
small and $n(S)$ has a steep decrease. Approaching $\alpha_c$,
the cluster size distribution $n(S)$ tends to a power law 
\beq{
n(S) \sim S^{-\tau},
}
where the value of the exponent was obtained as $\tau=2.35\pm 0.08$
which is higher than the corresponding exponent 
of 2d-percolation on a square lattice $\tau =
187/91 \approx 2.0549$ \cite{stauffer_percolation}. Note that in the
regime where spanning clusters exist 
($\alpha \geq 0.4$), the distribution $n(S)$ contains only the finite
clusters. 

\begin{figure}[h] 
  \begin{center} 
\epsfig{bbllx=100,bblly=350,bburx=430,bbury=640,file=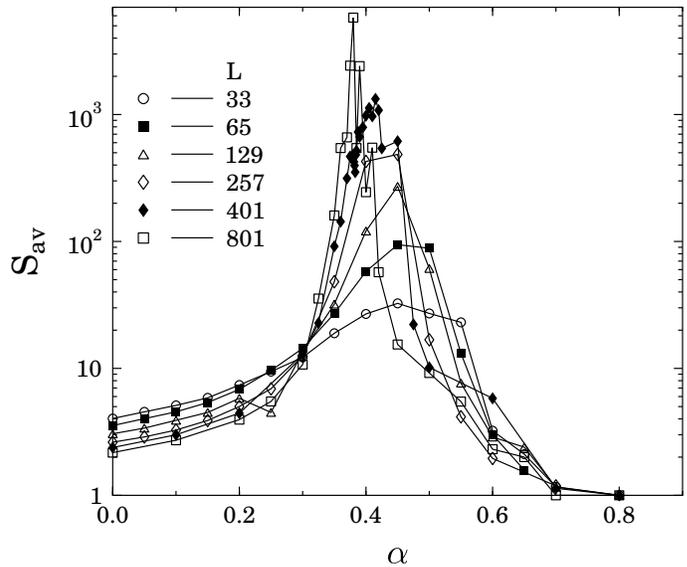,
  width=9.0cm}
   \caption{Average cluster size $S_{av}=m_2/m_1$ as a function of 
$\AL$ for different system sizes $L$. Simulation results were obtained
with a Weibull distribution $m=2$. }   \label{gr:cl_mass}  
  \end{center}
\end{figure}
\begin{figure}
  \begin{center}
\epsfig{bbllx=15,bblly=10,bburx=370,bbury=320,file=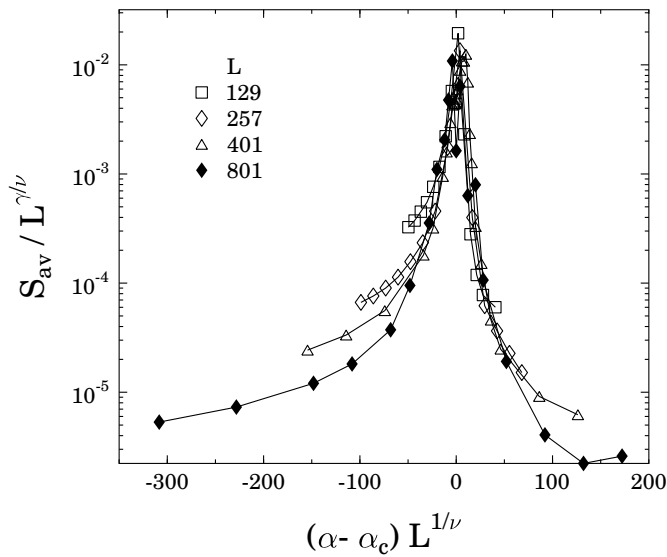,
  width=9.0cm}
   \caption{Finite size scaling of the average cluster size $S_{av}$
presented in Fig.\ \ref{gr:cl_mass}. The good quality collapse
obtained enables us to determine the value of the critical exponents
$\gamma$ and $\nu$ with a relatively good precision.
}   \label{gr:fss_sav} 
  \end{center}
\end{figure}

In order to characterize the evolution of the cluster structure when
$\AL$ is changed and to reveal the nature of the transition occurring
at $\alpha_c$, we calculated the average cluster size $S_{av}$
as the ratio of the second and first moments of the cluster
size distribution
\beq{
S_{av} = \frac{m_2}{m_1}.
}
The $k$-th moment $m_k$ of the distribution $n(S)$ is defined as
\beq{
  m_k = \sum \limits_{S}  S^k n(S) -S_{max}^k,
 }
where the largest cluster is excluded from the summation. 
Figure \ref{gr:cl_mass} presents $S_{av}$ as a function of $\AL$ for
different system sizes ranging from $L=33$ to $L=801$. 
It can be seen in the figure that for each value of $L$ the average
cluster size $S_{av}$
has a maximum at a well defined value of $\alpha$, which becomes a
sharp peak with increasing $L$, {\it i.e.} the peak becomes higher and
narrower for larger systems. The observed behavior is typical for
continuous phase transitions, where the position of the maximum defines
the critical point of the finite size system. Based on the analogy to
critical phenomena we tested the validity of the scaling law
$S_{av}\sim L^{\gamma/\nu}\phi((\alpha-\alpha_c)L^{1/\nu})$, where
$\phi$ denotes the scaling function of $S_{av}$
\cite{newman_barkema_1999,stauffer_percolation}. The results presented in Fig.\ 
\ref{gr:fss_sav} were obtained by varying the values of the critical
point $\alpha_c$ and of the critical exponent of the susceptibility $\gamma$,
and correlation length $\nu$ until the best data collapse
was reached.  
It can be observed in Fig.\ \ref{gr:fss_sav} that in the vicinity of
the critical point $\alpha_c$ a good quality data collapse is obtained
using the values $\alpha_c = 0.385 \pm 0.01$, $\gamma = 2.0 \pm 0.15$,
and $\nu = 1.0 \pm 0.1 $, where the critical exponents are only
slightly different from the percolation exponents of $\gamma = 43/18 \approx 2.389$
and $\nu =4/3 \approx 1.33 $ in 2d
\cite{stauffer_percolation}.   

\begin{figure}
\begin{center}
\epsfig{bbllx=80,bblly=330,bburx=440,bbury=640,file=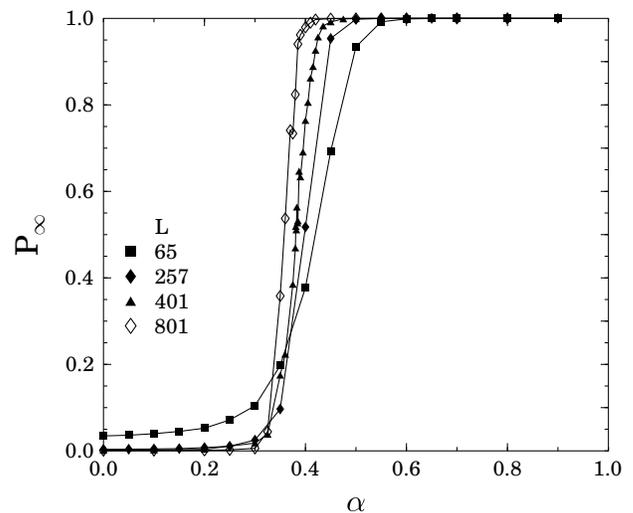,
  width=8.7cm} 
\caption{Order parameter $P_{\infty}$ vs.\ $\AL$ for several system
sizes $L$ with the Weibull index $m=2$. 
}  
\label{gr:pinf:a}     
\end{center}
\end{figure}

\begin{figure}
\begin{center}
\epsfig{bbllx=20,bblly=10,bburx=370,bbury=320,file=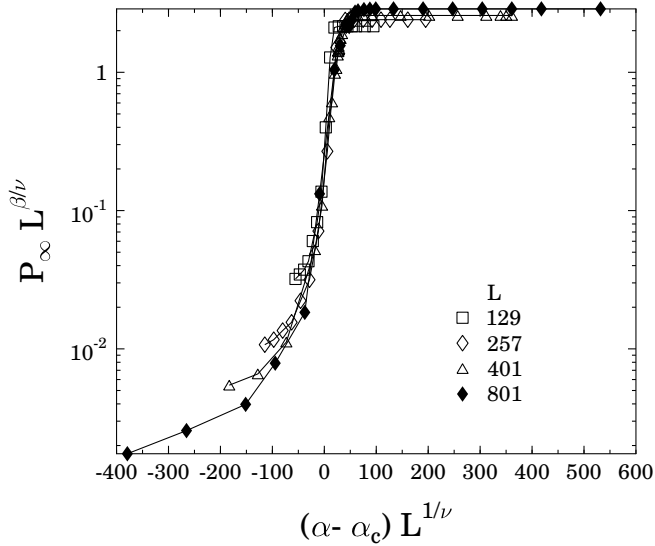,
  width=8.7cm}    
    \caption{Finite size scaling of the order parameter $P_{\infty}$
presented in Fig.\ \ref{gr:pinf:a}. The parameter values used to
obtain the best quality data collapse agree well with the ones
determined by the finite size scaling of $S_{av}$.} 
\label{gr:fss_pinf} 
  \end{center}
\end{figure}

At the critical point a spanning cluster of broken fibers
occurs which is much larger than the other clusters. In order to
characterize the strength of the spanning cluster we calculated the
probability $P_{\infty} (\AL)$ that a failed fiber belongs 
to the largest cluster. For percolation the
quantity $P_{\infty}$ plays the role of the order parameter whose
value distinguishes the phases of the system.
Similarly to percolation lattices, we find numerically a sharp rise
from $P_{\infty} =0$ to $P_{\infty} =1$  at $\AL _c \approx 0.4$, see
Fig.~\ref{gr:pinf:a}. When the system size $L$ is increased 
$P_{\infty}$ tends to a step function indicating that the transition
becomes sharper. Assuming the scaling law $P_{\infty} \sim
L^{-\beta/\nu}\psi((\alpha-\alpha_c)L^{1/\nu})$ of the order
parameter for finite size systems, where $\psi$ denotes the scaling
function and $\beta$ is the order parameter exponent
\cite{newman_barkema_1999,stauffer_percolation}, we replotted the 
data in Fig.\ \ref{gr:fss_pinf}. The good quality of the data collapse
was obtained with the parameter values $\alpha_c =0.33 \pm 0.01$,
$\beta = 0.15 \pm 0.06$, and
$\nu = 0.95 \pm 0.1$. Note that the value of $\nu$ agrees well with the one
determined by the finite size scaling of the average cluster size
$S_{av}$, larger deviations occur only for the critical point
$\alpha_c$. The order parameter exponent $\beta$ is compatible with the percolation value $\beta = 5/36 \approx 0.13$ 

\subsection{Random crack nucleation versus crack growth}

The failure mechanism of disordered materials and its relation to the
amount of disorder has long been discussed
in the literature
\cite{hh_smfdm,chakrab_beng_book_1997,kun_epjb_2000,zapperi_percol_localiz_2004,zapperi_pre59_1999,zapperi_prl78_1997,hansen_jpfrance_1989,batrouni_hansen_fuseprl_1998}.
When the material has a low degree of 
disorder only a small amount (if any) of damage occurs prior to macroscopic
failure. In this case even the nucleation of a single microcrack can lead to
localization and abrupt failure of the system. Increasing the amount
of disorder, the macroscopic failure is preceded by a larger and larger
precursory activity, {\it i.e.} a large amount of damage accumulates
and local breakings can trigger bursts of breaking events
\cite{kloster_pre_1997}.
Since cracks nucleate randomly, the process of damage before
localization resembles percolation up to some extent. Stress
concentration builds up around failed regions which might lead to correlated
growth of the nucleated cracks
\cite{ray_physa229_1996,kun_epjb_2000,zapperi_percol_localiz_2004,andersen_tricrup_prl_1997}.
Increasing the strength of disorder, correlation effects 
become less dominating and in the limit of infinite disorder the
damage accumulation process can be mapped to percolation
\cite{roux_jsp52_237_1988}.    

\begin{figure}
  \begin{center}
\epsfig{bbllx=140,bblly=330,bburx=490,bbury=620,file=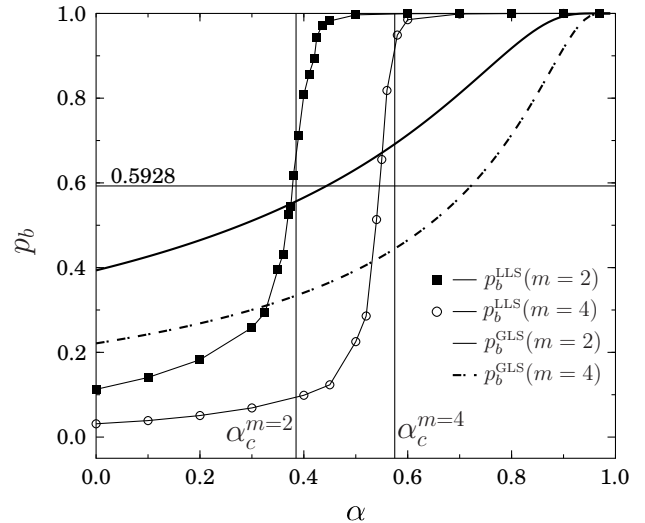,
  width=8.7cm}
   \caption{
The fraction of broken fibers $p_b$ at $\sigma_c$ as a function of $\alpha$ for
fiber bundles of LLS and GLS with different strength of disorder $m=2$
and $m=4$. The
vertical line indicates the critical point obtained as 
the position of the maximum of the average cluster size (see Fig.\
\ref{gr:cl_mass}). The critical 
probability of percolation $p_c$ on the square lattice is indicated by
the horizontal line. Note that for both disorder distributions in LLS,
the location where $p_b^{LLS}$ exceeds $p_c$ practically coincides with the
corresponding critical point $\alpha_c$, indicating the percolation
nature of the transition.
 }   \label{fig:dens_broken} 
  \end{center}
\end{figure}

We have shown above that in the plastic fiber bundle model (PFBM), the load
bearing capacity of broken fibers 
has a substantial effect on the process of failure when the load
redistribution is localized
due to the reduction of the stress concentration along cracks.
In order to give a quantitative characterization of damage
accumulation in our model, we determined the fraction of broken fibers
$p_b$ at global failure $\sigma_c$ as a function of the strength of plasticity
$\alpha$.  
The quantity $p_b$ can also be interpreted as the probability $0 \leq
p_b \leq 1$ that a randomly chosen fiber in the bundle is broken
which makes it possible to compare the spatial structure of damage to
percolation lattices \cite{stauffer_percolation} generated with the
occupation probability $p=p_b$
\cite{hh_smfdm,chakrab_beng_book_1997,raul_varint_2002,roux_jsp52_237_1988}.
The results are presented in Fig.\ \ref{fig:dens_broken} for the
system size $L=401$ and Weibull parameters $m=2$ and $m=4$
plotting also the corresponding GLS results for comparison. 
In the case of local load sharing, when
the failure load of fibers is almost entirely redistributed locally ($\alpha
\approx 0$) only a small damage can accumulate up to global failure
$p_b^{LLS}\approx 
0.1-0.2$ keeping the integrity of the system. Comparing the curves of different
Weibull indices $m$ it follows that the stronger the disorder is, the
larger amount of damage the system can tolerate at the same value of
$\alpha$. In the vicinity of the respective $\alpha_c$, the breaking fraction
$p_b^{LLS}$ rapidly  increases and converges to the maximum value
$p_b^{LLS} \approx 1$, which implies that in the regime  $\alpha > \alpha_c$
practically no localization occurs, the bundle can remain stable until
almost all fibers break. 

It is instructive to compare this behavior to the case of GLS, where
those fibers break up to the critical point whose breaking threshold
falls below $\sigma_c$, hence, $p_b^{GLS}(\alpha)$ can simply be obtained as
$p_b^{GLS}=P(\sigma_c(\alpha))$.
It can be seen in Fig.\ \ref{fig:dens_broken} that under global load
sharing for $\alpha \approx 0$ a significantly larger fraction of
fibers fails without destroying 
the system than in the LLS bundle. 
The breaking fraction $p_b$ is a monotonically increasing function of
$\alpha$ irrespective of the range of load sharing, however, in the
vicinity of the critical point of LLS bundles $p_b^{LLS}$ exceeds the
smoothly rising GLS curves $p_b^{GLS}$. 
Note that depending on the threshold distribution $P$ of fibers, even
at $\alpha =0$ the value of $p_b^{GLS}$ can be
smaller or larger than the critical percolation probability $p_c$ of
the corresponding lattice type, since (contrary to fuse networks
\cite{batrouni_hansen_fuseprl_1998,zapperi_percol_localiz_2004} or
discrete element models \cite{HJH_cont_and_discont_2000})
fracture in fiber bundles is not  
related to the appearance of a spanning cluster of failed elements.
Varying $\alpha$ as a control parameter,
formally the GLS results could be perfectly mapped onto a percolation
problem: at the critical value of the control parameter
$\alpha_c^{GLS}$ defined as $P(\SC(\AL_c^{GLS})) = p_c$ a spanning
cluster occurs, which has 
a fractal structure, the average size of finite clusters has a maximum
at the critical point and the cluster size distribution exhibits gap
scaling \cite{stauffer_percolation}. However, this percolation is not
related to the point of failure of the GLS bundle, the analogy to
percolation is based purely on geometrical properties without any
physical relevance.  

Figure \ref{fig:dens_broken} shows that for localized load
sharing the phase transition occurs when the damage fraction
$p_b^{LLS}$ reaches the critical percolation probability $p_c$ of the
corresponding lattice type. 
Due to the very localized load sharing, only
short range correlations arise in the system which are further
moderated by the finite load bearing capacity of broken fibers. Hence,
in the vicinity of the transition point $p_b^{LLS}(\alpha_c) \approx
p_c$ holds and the evolution of the microstructure of damage shows
strong analogy to percolation lattices. It can be seen in Table
\ref{tab:table1} that the critical exponents of the plastic fiber
bundle model are slightly different from the corresponding exponents  
of percolation, furthermore, the usual scaling relations of
percolation critical exponents \cite{stauffer_percolation} are not
fulfilled within the error bars. It has been shown for percolation
that correlated occupation probabilities lead to the same critical
behavior as random percolation when the correlations are short ranged 
\cite{weinrib_pra_1984,family_pra_1988}, however, long range
correlations result in changes of the critical exponents
\cite{weinrib_pra_1984}. It is interesting to note that the value of
the correlation length exponent $\nu$ of PFBM is smaller
than the value of random percolation which is consistent with the
presence of relevant correlations \cite{weinrib_pra_1984}.
We would like to emphasize that
contrary to global load sharing, this percolation like transition has
important physical consequences on the behavior of the fiber
bundle. The failure process of the bundle is dominated by the
competition of fiber breaking by local stress enhancement due to load
redistribution and by local weakness due to disorder. Our detailed
analysis revealed that the relative
importance of the two effects is controlled by the parameter
$\alpha$. Below the critical point $\alpha < \alpha_c$ high stress
concentration can develop around cracks so that the failure of the
bundle occurs due to localization. 
Above the critical point $\alpha \geq \alpha_c$ the macroscopic
response of the LLS bundle becomes practically identical with the GLS
constitutive behavior showing the 
dominance of disorder. It is important to note that the size
distribution of bursts of simultaneously failing fibers becomes a
power law at the critical point $\alpha_c$ with an exponent $\mu$
equal to the value recently predicted
for GLS bundles of so-called critical failure threshold distributions
\cite{hansen_crossover_prl,hansen_lower_cutoff_2005}. This can be
explained such that the large avalanches of power law distribution
occurring in the plastic fiber bundle model at $\alpha_c$ (see Fig.\
\ref{gr:lls_avals}) are dominated by the strong fibers of the bundle
whose strength distribution is close to critical
\cite{hansen_crossover_prl,hansen_lower_cutoff_2005}. 

\begin{figure}
  \begin{center}
\epsfig{file=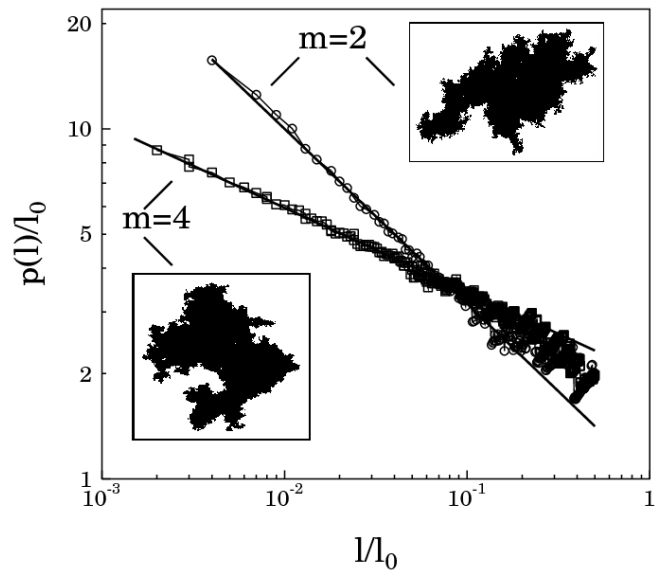,
  width=8.7cm}
   \caption{
Structure of the spanning cluster at two different disorder strengths in a
lattice of size $L=401$. The
perimeter length $p(l)$ of the cluster is plotted as function of the
length $l$ of the yardstick normalized by the side length $l_{0}$ of
the inscribing square. The insets present the clusters analyzed. 
 }   \label{fig:fractal_cluster} 
  \end{center}
\end{figure}
The structure of the spanning cluster of the LLS bundle formed at the
critical point $\alpha_c$ has also remarkable features different from the
spanning cluster of percolation \cite{stauffer_percolation}. The insets
of Fig.\ \ref{fig:fractal_cluster} present representative examples
of the spanning cluster of a system of size $L=401$ at two different
disorder strengths. It can be
observed that the clusters are compact, they practically do not have
holes, there are no islands of unbroken fibers in the interior of the
cluster. This structure is a direct consequence of the merging of
growing compact clusters where espcially large stress concentrations
arise between the cluster surfaces breaking the fibers and filling the
holes in the spanning cluster. We note that in the limiting case of
very strong disorder a small amount of intact fibers may
survive dispersed over the spanning cluster. The result implies that
the fractal dimension of the  
spanning cluster of the LLS bundle is 2, which should be compared to
the corresponding value of random percolation $D=91/48 \approx 1.896$ where a
finite amount of holes exists \cite{stauffer_percolation} even for
short range correlated occupation probabilities
\cite{family_pra_1988}. 
The perimeter of the spanning cluster, however, has a fine structure,
{\it i.e.} it has a large number of peninsulas and valleys of all sizes. To reveal the 
structure of the perimeter, we measured its length $p(l)$ as a function of
the length of the yardstick $l$. It can be seen in Fig.\
\ref{fig:fractal_cluster} that $p(l)$ shows a power law dependence on
$l$ over almost two decades
\beq{
p(l) \sim l^{-\delta_p},
\label{eq:fractal}
}
where the value of the exponent proved to be $\delta_p = 0.5\pm 0.03$
for a Weibull distribution of fiber strength with $m=2$.
The power law Eq.\ (\ref{eq:fractal}) indicates that the perimeter
line is a fractal with a dimension $D_p = 1+\delta_p =1.5\pm0.03$. The
upper bound of the scaling range in Fig.\ \ref{fig:fractal_cluster}
can be attributed to the characteristic size of peninsulas of the
spanning cluster, over which the rough structure of the perimeter
disappears. Numerical calculations revealed that the fractal dimension
of the cluster surface $D_p$ is not universal, {\it i.e.} it depends
on the strength of disorder of the breaking thresholds. The insets of Fig.\
\ref{fig:fractal_cluster} illustrate that a lower amount of disorder
gives rise to a more regular, smoother cluster surface characterized
by a lower value of $D_p$. For the Weibull index $m=4$ we obtained
$D_p = 1.24 \pm 0.05$, which is significantly smaller than the
corresponding value of $m=2$. The surface of damage clusters should be
compared to the hull of the spanning cluster of percolation with the
fractal dimension $D_p=7/4 = 1.75$ \cite{saleur_prl_1987} (see also
Table \ref{tab:table1}).     
\begin{table}
\begin{center}
\caption{ \label{tab:table1} Summary of the critical exponents of the
plastic fiber bundle model with local load sharing. For comparison the
value of the corresponding critical exponents of percolation are also
shown. For the perimeter fractal dimension $D_p$ of PFBM a range is given.}
\begin{tabular}{|c|c|c|}
\hline
Critical exponents & PFBM           &  Percolation        \\
\hline
$\beta$         & $0.15 \pm 0.06$ &  $5/36 \approx 0.13$   \\
\hline
$\gamma$        & $2.0\pm 0.15$   &  $43/18\approx 2.38$   \\
\hline
$\tau$          & $2.35\pm 0.08$  &  $187/91\approx 2.05$  \\
\hline
$\nu$           & $1.0 \pm 0.1$   &  $4/3 \approx 1.33$    \\ 
\hline
$D$             & $2.0$           &  $D=91/48 \approx 1.896$ \\
\hline
$D_p$           & $1.0-2.0$       &  $7/4=1.75$            \\
\hline
$\mu$ (Bursts)  & $1.5\pm 0.07$   &  --                    \\ 
\hline   
\end{tabular}
\end{center}
\end{table}

\section{Summary}
We introduced a fiber bundle model where failed fibers retain a
fraction $0 \leq \alpha \leq 1$ of their failure load. The value of
the parameter $\alpha$ interpolates between the perfectly rigid
failure $\alpha = 0$ and perfect plasticity $\alpha = 1$ of fibers.  
We carried out a detailed study of the effect of the finite load
bearing capacity of fibers on the microscopic damage process and
macroscopic response of fiber bundles considering both global and
local load sharing for the load redistribution after fiber failure.   
Analytic calculations and computer simulations revealed that under
global load sharing the macroscopic constitutive behavior of the
interface shows a transition to perfect plasticity when $\alpha
\rightarrow 1$, where the yield stress proved to be the average fiber
strength. Approaching the state of perfect plasticity, the size
distribution of bursts has a crossover from the mean field power law
form of exponent 2.5 to a faster exponential decay.

When the 
load sharing is localized it is found that the load carried by the
broken fibers has a stabilizing effect on the bundle, {\it i.e.}
it lowers the stress concentration around clusters of failed fibers
which has important consequences on the microscopic process of
fracture and on the macroscopic response of the bundle. 
Extensive numerical calculations showed that at a specific value
$\alpha_c$ a very interesting transition occurs from a phase where
macroscopic failure 
emerges due to stress enhancement around failed regions leading to
localization, to another phase where the disordered fiber strength plays
the dominating role in the damage process. 

On the macro-level, below the critical point $\alpha < \alpha_c$ the
fiber bundle shows a brittle response, {\it i.e.} the macroscopic
failure is preceded by a weak non-linearity, while for $\alpha \geq
\alpha_c$ the constitutive behavior of the LLS bundle becomes
practically identical with the GLS counterpart. 
Analyzing the evolution of
the micro-structure of damage with increasing $\alpha$, the transition
proved to be continuous analogous to percolation. Computer
simulations revealed that the avalanche
size distribution of fiber breakings becomes a power law at the critical
point with an universal exponent equal to the mean field exponent of bundles
with critical strength distributions. The spanning cluster of failed
fibers formed at the transition point proved to be compact with a
fractal boundary whose dimension increases 
with the amount of disorder. The critical value $\alpha_c$ is not
universal, besides the lattice structure, it also depends on the
strength of disorder.

The plastic fiber bundle model can be relevant for the shear failure
of interfaces where failed surface elements can remain in contact
still transmitting load. Such glued interfaces of solids typically
occur in fiber composites, where fibers are embedded in a matrix
material. The finite load bearing capacity of failed elements of the
model can account for the frictional contact of debonded fiber-matrix
interfaces and also for plastic behavior of the components.

\begin{acknowledgments}
This work was
supported by the Collaborative Research Center SFB 381. F.\ Kun
acknowledges financial support of the Research Contracts
NKFP 3A-043/2004, OTKA M041537, T049209 and of the Gy\"orgy B\'ek\'esy
Foundation of the Hungarian Academy of Sciences.
\end{acknowledgments}

\bibliography{raischel}
\end{document}